\documentclass[fleqn,12pt,twoside]{article}
\usepackage{espcrc1}

\usepackage{graphicx}

\usepackage[figuresright]{rotating}

\newcommand{\AmS}{{\protect\the\textfont2
  A\kern-.1667em\lower.5ex\hbox{M}\kern-.125emS}}

\title{ Analysis of the characteristics of nucleus-nucleus collisions 
depending on the centrality }

\author{ M.K. Suleymanov \address[MCSD]{ Laboratory of High Energies, JINR, \\ 
        Dubna, Moscow region, 141980, Russia}%
        \thanks{E-mail: mais@sunhe.jinr.ru},
         O.B. Abdinnov\address[MCSD1]{ Physics Institute of AS  \\ 
        Baku, Azerbaijan},
	N.S. Angelov\address[MCSD2]{ Laboratory of Nuclear Problems, JINR, \\ 
        Dubna, Moscow region, 141980, Russia},
        B.Z. Belashev\address[MCSD3]{Institute of Geology, Karelian RC, RAS,\\
        Petrozavodsk, Karelia, Russia},
	Ya.G.Guseynaliyev\address[MCSD4]{ AMAA of Azerbaijan  \\ 
        Baku, Azerbaijan},
        A.A. Kuznetchov\addressmark[MCSD], A.S.Vodopianov\addressmark[MCSD]	}
       
\begin{document}

\maketitle

\begin{abstract}

The experimental results on some centrality depending characteristics   of hadron-nuclear and nuclear-nuclear interactions at high energies demonstrate the regime changes. Appearance of  strong interaction matter's mixed states is considered as a cause of  it and the effect of a cluster formation is discussed as  one of  the phenomena connected with the mixed states. 

\end{abstract}

\section{INTRODUCTION}

One of  the important experimental methods to get the information on the changes
of states of nuclear matter with increasing its baryon density is to study the
characteristics of hadron-nuclear and nuclear-nuclear interactions depending on
the centrality of collisions ($Q$) at high energies. There are some results
obtained in these experiments for the interactions of $\pi$-mesons, protons and
nuclei  with nuclei at energies less than   SPS' energies which demonstrate
existence of the regime changes in these dependencies~\cite{[1]}-\cite{[5]}.  
It is necessary to note that in  different experiments the values of $Q$ are 
defined in  different ways. Therefore it is very difficult to compare the 
obtaineed results on $Q$-dependencies in  different papers . 

QCD predicts that at high energy density, hadronic matter will turn into 
plasma of deconfined quarks and gluons~\cite{[7]}. It is expected  that  the 
temperature of hadron matter T  will be $T > T_c \simeq 150-200 MeV$ and $\mu_{B}$ 
will be $\mu_{B}> \mu_{B_c}$ ($\mu_{B}$ increases with the barion charge). It is a new phase of 
nuclear matter. The $T_c$ could be reached at energies of SPS, RICH and LHC. 

To explain the above mentioned results we consider the possibility of the 
phase transition at $ T < T_c $ in the system with high barion density (at high $\mu_B$). 
In such systems the neighbouring nucleons could form the cluster and  neighbouring quarks 
could form the diquarks ( for example in the result of percolation). So these systems are 
a mixed system (MS) of compressed nucleons (clusters) and diquarks which could appear at  energies 
of GSI, Synchrophasotron  (Nuclatron) and AGS.

Experimental information  on the conditions of the MS appearance could give 
the  possibilities to fix  the onset of the deconfinement. It is
 important for further  separate on of  the   effects connected with 
 deconfinement of strong   interaction matter   from the other ones. 

The regime changes  which were shown above could be a better indication of 
the MS appearance in  high energy interactions, but they are not enough  to 
assert it. For further confirmation of the appearance and existence of the MS 
it  is necessary to obtain an additional experimental information because 
the regime changes under consideration can be explained by some other 
approaches without the MS.

Let us discuss  experimental possibilities to get a signal on the MS. First 
of all we have to answer  a question, by what experimental observable 
effects the MS could be accompanied. It is  clear that the first effect is 
a cluster formation ( for example in the result of percolation), the second 
effect could be the appearance of meson condensation (which could be formed 
in the result of  hadronisation of diquarks). 
                        
There are many papers in which the processes of nuclear 
fragmentation~\cite{[9]} and the processes of central 
collisions~\cite{[10]} are considered as  critical phenomena and 
percolation approach is suggested  to be used to explain these phenomena. 
We have used some ideas from these works to experimental search of a signal 
on a cluster formation. We suppose that in  hadron-nucleus and  nucleus-nucleus 
collisions the cluster could appear on some critical values of $Q$ and 
would decay into  fragments and free nucleons. The number of clusters and  
fragments would increase with $Q$ in the interval less than the critical values 
of $Q$ (for MS formation) and then their values would decrease with the increase 
of $Q$ to the boundary of the central collisions region. It could lead to the regime 
change in the behavior 
of different characteristics of events  depending on $Q$ and the number of fragments. 
We believed that if  the cluster  exists   and if it is  a source of fragments then 
the influences of nuclear fragments on the behaviour of the events' characteristics 
depending on $Q$  could have a critical character.

\subsection{EXPERIMENT}

To test this idea  the behaviors of  the events' number  depending on $Q$ have been 
studied by us ~\cite{[20]}. The values of $Q$  were determined  in two variants.  
In the first variant the values of $Q$ were determined as a number of protons 
emitted in a event and in the second variant -- as a number of protons and 
fragments emitted in the events. We have used 20407 $^{12}CC$  events  at the 
momentum of  4.2 A GeV/c~\cite{[11]}. The experimental data were compared 
with the simulation data coming from the  quark-gluon string model (QGSM) 
without nuclear fragments~\cite{[12]}.  We want to note that  the behavior of 
the events' number depending on $Q$ determined for both variants have to be
 similar if there are no  cluster and they would differ if the cluster exists 
 as a source of fragments. The distributions of the events' number as a function of 
 $Q$ are shown in fig.  1a,b . The open symbols correspond to the first 
 variant of $Q$-determination, the black symbols correspond to the second 
 ones( the fragments were included). It is seen that for the cases when the 
 fragments numbers were included to determine $Q$ the form of the distributions 
 sharply changes and has  two steps structure (black symbols in fig. 1a). 
The $Q$-dependencies of the events' number coming from the QGSM are shown 
in fig. 1b . The open symbols correspond to the cases without  stripping 
protons  and the black symbols correspond to the cases with the stripping 
protons. It is seen that the form of the distribution strongly differs from 
the experimental one  in fig. 1a. There is no two steps structure in this 
figure. Therefore  we could assure that the observed difference is connected 
with the fragment influence. 
This result demonstrates that the influence of nuclear fragmentation processes 
in the behavior of the events' number depending on $Q$  has a critical 
character. 
To explain this result we can suppose that it could be connected with the 
existence of a cluster. It is possible that with the increase of $Q$ the 
probability of cluster formation grows but further increasing of the $Q$ 
( in the region of high $Q$) leads to a cluster decay on nuclear fragments 
and then on free nucleons. It could be the reason of the  observed  
two step structure in the distributions. The first step connected with the 
formation of  a cluster and the second one with its decay. The obtained 
results on multifragment production at high energy nucleus-nucleus interaction 
could give additional confirmations for it (for example see ~\cite{[13]}). 

\begin{figure}[htb]
\includegraphics[height=15pc,width=30pc]{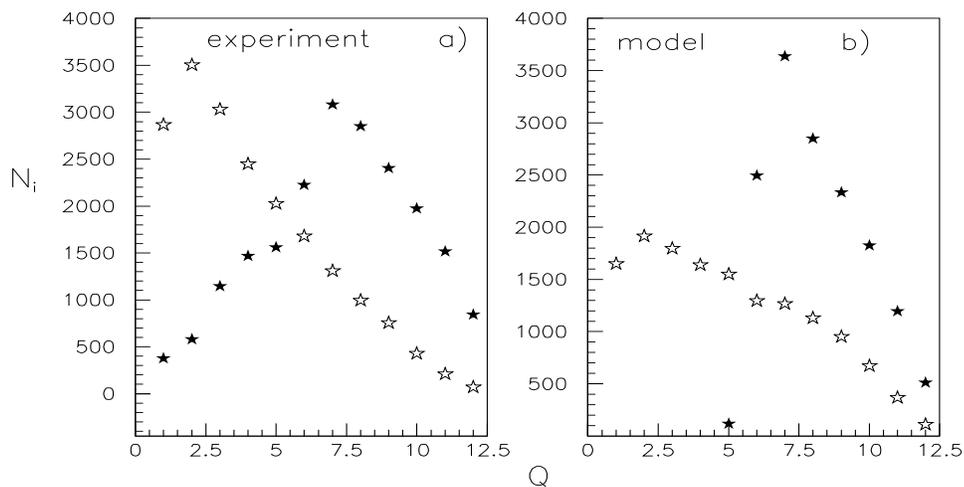}
\caption{The distributions of events' number as a function of $Q$ for the $^{12}CC$ events at the momentum of 4.2 A GeV/c ; a)the experimental data b) 
	the simulation data. }
\label{fig:1}
\hspace{\fill}
\end{figure}


\begin{thebibliography}{9}
\bibitem{[1]} O.B. Abdinov et  al. JINR Rapid communications 
               No 1[75]-96 (1996) 51.
\bibitem{[2]} S.Vokal,M.Sumbera.Yad.Fiz.39 (1984) 1474. 
\bibitem{[3]} C.A. Ogilvie.J.Phys.G.Nucl.Part.Phys.25(1999)159.
\bibitem{[4]} M.I. Tretyakova. EMU-01 Collaboration. Proceeding of the 
              Xith International Seminar on High  Energy  Physics Problems. 
              Dubna, JINR, (1994)616. 
\bibitem{[5]} A.Abdelsalam et al. JINR E1-82-509, Dubna (1982).
\bibitem{[7]} H. Satz.  Nucl. Phys. A661(1999) 104.
\bibitem{[9]} J. Desbois, Nucl. Phys. A466 (1987) 724 ; J. Nemeth et al. 
              Z.Phys.A 325 (1986) 347; S. Leray et al. Nucl. Phys.A511(1990)414;
              ; A.J. Santiago and K.C. Chung J. Phys. G:Nucl.Part. Phys. 
              16 (1990)1483; G.Musulmanbekov, A.Al-Haidary. Russian    
              J.Nuc.Phys.,v.66 (2003) 1.
\bibitem{[10]}X.Campi, J. Desbois Proc. 23 Int. Winter Meeting on Nucl. 
              Phys. Bormio,1985; Bauer W. et al. Nucl. Phys. 452(1986) 699; 
              A.S. Botvina, L.V.Lanin. Sov. J. Nucl. Phys. 55 (1992) 381.
\bibitem{[20]} Suleymanov et al. Proceedings of the Conference: Bologna2000, Bologna, Italy, (2000)375.
\bibitem{[11]} N.Akhababian et al.- JINR Preprint 1-12114, Dubna, 1979.; 
               N.S.Angelov et al.- JINR  Preprint 1-12424, Dubna, 1989 ; 
               A.I.Bondorenko  et al., JINR Communication, P1-98-292, Dubna, 
               1998; M. K.Suleimanov  et al. Phys.Rev.C. 58 (1998)351.
\bibitem{[12]} N.S. Amelin, L.V.Bravina, Sov. J. Nucl. Phys. 51 (1990) 211
               ; N.S. Amelin et al.,  Sov. J. Nucl.  Phys.50 (1990) 272.
\bibitem{[13]} W.Resisdorf. Dynamics of multufragmentation  in Heavy Ion Collisions. 
               E-print: nucl-ex/0004008, v.1 (2000).
                 

\end{thebibliography}
\end{document}